\documentclass[a4paper,12pt]{article}
\usepackage{epsfig}
\newcommand{\nc}{\newcommand}
\nc{\non}{\nonumber}
\nc{\pt}{p_{{}_T}}
\nc{\lmc}{\Lambda_c^+}
\nc{\plmc}{\vec{\Lambda}_c^+}
\nc{\all}{A_{LL}}
\nc{\dg}{\Delta G(x,Q^2)}
\nc{\stil}{\tilde{s}}
\nc{\ttil}{\tilde{t}}
\nc{\util}{\tilde{u}}
\nc{\shs}{\hat{s}}
\nc{\ths}{\hat{t}_1}
\nc{\uhs}{\hat{u}_1}
\nc{\cosec}{\rm cosec}
\nc{\mpr}{m_p}
\nc{\mc}{m_c}
\nc{\mlc}{m_{\lmc}}
\nc{\pa}{p_{{}_a}}
\nc{\pb}{p_{{}_b}}
\nc{\pA}{p_{{}_A}}
\nc{\pB}{p_{{}_B}}
\nc{\plc}{p_{\lmc}}
\nc{\pc}{p_{{}_c}}
\def\gesim{\lower0.5ex\hbox{$\:\buildrel >\over\sim\:$}} 
\def\lesim{\lower0.5ex\hbox{$\:\buildrel <\over\sim\:$}} 
\nc{\prd}[3]{{\rm Phys.\ Rev.}\ {{\bf D{#1}} (#2), #3}}
\nc{\prl}[3]{{\rm Phys.\ Rev.\ Lett.}\ {{\bf {#1}} (#2), #3}}
\nc{\plb}[3]{{\rm Phys.\ Lett.}\ {{\bf B{#1}} (#2), #3}}
\nc{\npb}[3]{{\rm Nucl.\ Phys.}\ {{\bf B{#1}} (#2), #3}}
\nc{\ptp}[3]{{\rm Prog.\ Theor.\ Phys.}\ {{\bf {#1}} (#2), #3}}
\nc{\zfp}[3]{{\rm Z.\ Phys.}\ {{\bf C{#1}} (#2), #3}}
\nc{\mpla}[3]{{\rm Mod.\ Phys.\ Lett.}\ {{\bf A{#1}} (#2), #3}}
\nc{\rmp}[3]{{\rm Rev.\ Mod.\ Phys.}\ {{\bf {#1}} (#2), #3}}
\nc{\ijmpa}[3]{{\rm Int.\ J.\ of\ Mod.\ Phys.}\
               {{\bf A{#1}} (#2), #3}}
\nc{\epj}[3]{{\rm Eur.\ Phys.\ Jour.}\ {{\bf C{#1}} (#2), #3 }}
\begin{document}
\pagestyle{empty} \setlength{\footskip}{2.0cm}
\setlength{\oddsidemargin}{0.5cm} \setlength{\evensidemargin}{0.5cm}
\renewcommand{\thepage}{-- \arabic{page} --}
\def\mib#1{\mbox{\boldmath $#1$}}
\def\bra#1{\langle #1 |}      \def\ket#1{|#1\rangle}
\def\vev#1{\langle #1\rangle} \def\dps{\displaystyle}
\def\sla#1{\mbox{$#1\hspace*{-0.17cm}\scriptstyle{/}\:$}}
\renewcommand{\thepage}{-- \arabic{page} --}
   \def\thebibliography#1{\centerline{REFERENCES}
     \list{[\arabic{enumi}]}{\settowidth\labelwidth{[#1]}\leftmargin
     \labelwidth\advance\leftmargin\labelsep\usecounter{enumi}}
     \def\newblock{\hskip .11em plus .33em minus -.07em}\sloppy
     \clubpenalty4000\widowpenalty4000\sfcode`\.=1000\relax}\let
     \endthebibliography=\endlist
   \def\sec#1{\addtocounter{section}{1}\section*{\hspace*{-0.72cm}
     \normalsize\bf\arabic{section}.$\;$#1}\vspace*{-0.3cm}}
\vspace*{-1.6cm}\noindent
\hspace*{11.cm}hep-ph/0009004\\

\vspace*{2cm}

\begin{center}
{\Large  \bf $\Lambda_c^+$ Production
    in Polarized $pp$ Scattering and Polarized Gluon Distribution}
\end{center}

\vspace*{1.5cm}
\begin{center}
\renewcommand{\thefootnote}{\alph{footnote})}
{\small \sc Kazumasa OHKUMA}$\,^{1),}$\footnote{E-mail address:
\tt ohkuma@radix.h.kobe-u.ac.jp},
{\small \sc Kazutaka SUDOH}$\,^{1),}$\footnote{E-mail address:
\tt sudou@radix.h.kobe-u.ac.jp} and
{\small \sc Toshiyuki  MORII}$\,^{1),2),}$\footnote{E-mail address:
\tt morii@kobe-u.ac.jp} 
\vspace{0.5cm}

$1)$ {\sl Graduate School of Science and Technology,
Kobe University}\\ {\sl Nada, Kobe 657-8501, JAPAN}\\
\vspace{0.3cm}
$2)$ {\sl Faculty of Human Development,
Kobe University}\\ {\sl   Nada, Kobe 657-8501, JAPAN}\\

\end{center}

\vspace*{2.5cm}
\centerline{ABSTRACT}

\vspace*{0.4cm}

\baselineskip=20pt plus 0.1pt minus 0.1pt

To extract information about the polarized gluon distribution, $\dg$,
in the nucleons,
we propose $\lmc$ productions in polarized $pp$ scattering,
$p + \vec{p} \rightarrow \plmc + X $, 
which will be observed at RHIC experiment starting soon.
For this process, we have calculated the spin correlation differential
cross section, $d\Delta \sigma/d\pt$, and the spin correlation
asymmetry defined by
$ \all \equiv  [ d \Delta \sigma / d \pt ] / [ d \sigma / d\pt ]$.
We have found that the $\all$ is sensitive to the polarized gluon distribution
in the nucleon and thus the process is promising for testing 
$\Delta G(x,Q^2)$.\\
\vspace*{-0.5cm}
\\
PACS number(s): 13.88.+e, 14.20.Lg, 13.85.Ni
\\
$$
  To~appear ~in~ Phys. ~Lett. ~B
$$
\vfill
\newpage
\renewcommand{\thefootnote}{$\sharp$\arabic{footnote}}
\pagestyle{plain} \setcounter{footnote}{0} \setcounter{page}{1}

\baselineskip=21.0pt plus 0.2pt minus 0.1pt

The proton spin has long been considered to be given by the
sum of the constituent quark spin, $\Delta \Sigma$,
in the naive quark model, which  suggests 
$\Delta \Sigma =\Delta u + \Delta d = 1$ and $\Delta s =0$,
where $\Delta u$, $\Delta d$ and $\Delta s$ denote the amount of the 
proton spin carried by $u$, $d$ and $s$ quark, respectively.
However, the experimental data on polarized structure
function of nucleons, $g_1^{p(n)}(x,Q^2)$, have revealed that the contribution
of the constituent quark spin  to the proton spin is quite small,
$\Delta \Sigma \simeq 0.3$,
and the $s$ quark is negatively polarized with a rather large value
$\Delta s \simeq -0.12$~\cite{SMC}.
This has been called the ``proton spin puzzle''~\cite{psp}.
Actually, the proton spin is given by the sum of the spin of 
the quarks (valence and sea), gluons 
and the orbital angular momenta among them;
\begin{equation}
\frac{1}{2}=\frac{1}{2}\Delta \Sigma +\Delta G + \vev{L_Z}_{q+g} ,
\label{sum}
\end{equation}
where $\Delta G $ and $\vev{L_Z}_{q+g}$ represent the amount of the proton
spin carried by the gluon and the orbital angular momenta of quarks and gluons,
respectively.
To understand the physical ground of the sum rule of Eq.(\ref{sum}), 
it is important to know the behavior of the polarized partons 
in the nucleon.
However, knowledge about the polarized gluon density, $\dg$, in
a nucleon is still poor, 
though many processes have been proposed so far to
extract information about it.
On the other hand, the RHIC experiment which will start soon,
will provide us many useful data for 
extracting information about polarized gluons in the near future.

\begin{figure}[t]
\parbox[b]{0.46\textwidth}
{
  \begin{center}
    \epsfxsize=7cm
    \epsfbox{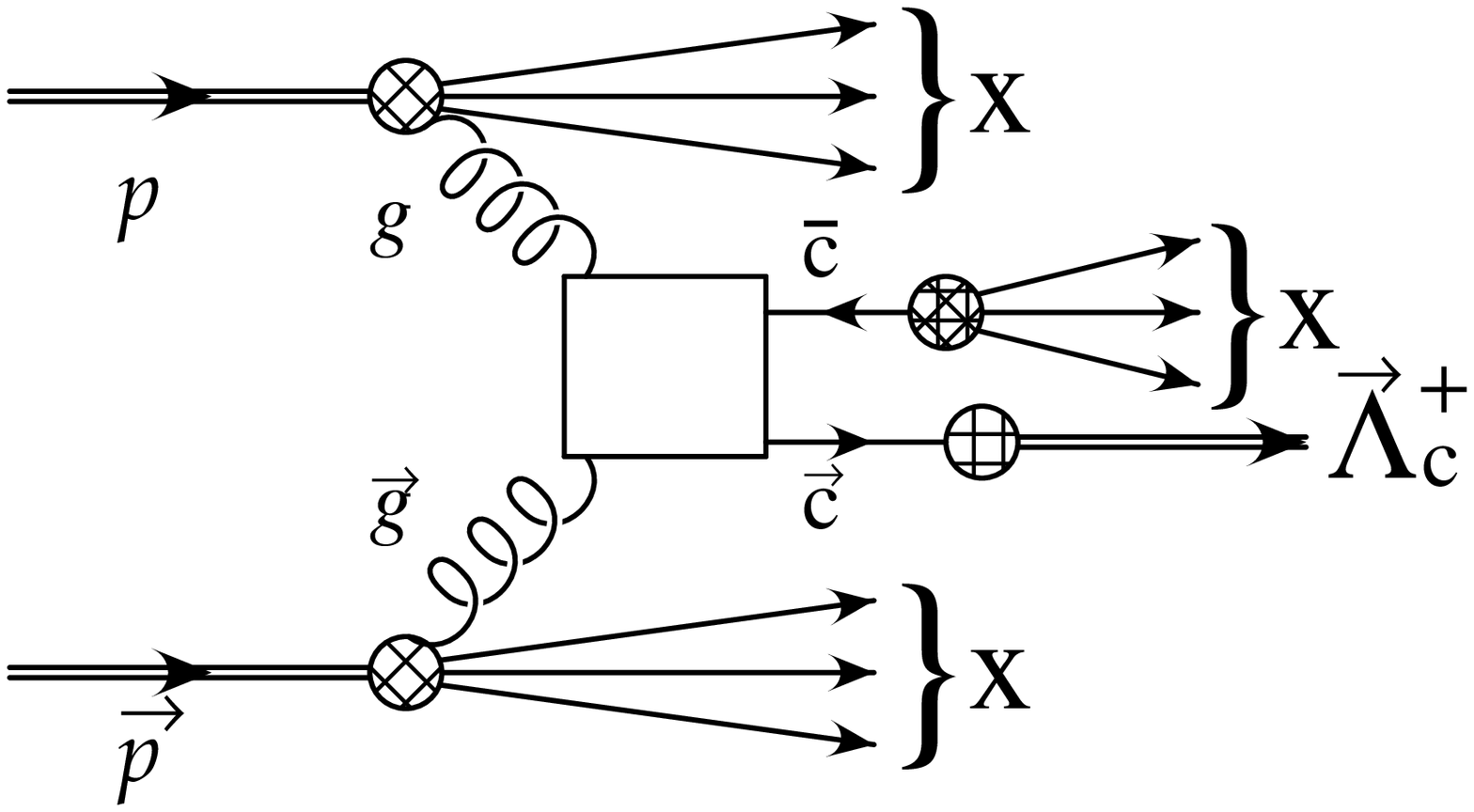}
  \end{center}
  \vspace*{-1cm}
  \caption{The diagram for
 $p + \vec{p}
 \rightarrow \plmc + X$ at the lowest order.} \label{mainpro}
} \hfill
\parbox[b]{0.46\textwidth}
{
  \begin{center}
    \epsfxsize=7cm
    \epsfbox{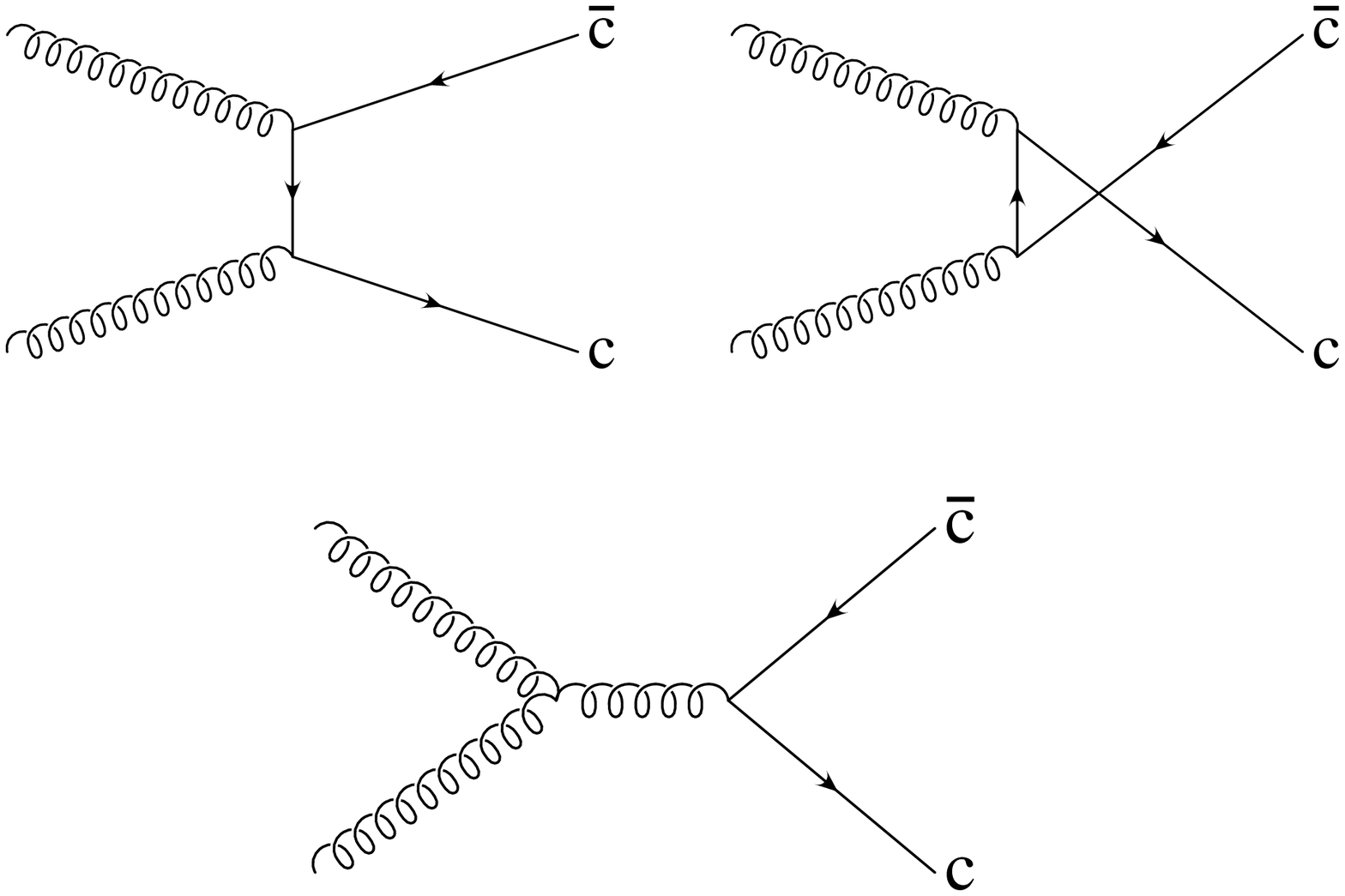}
  \end{center}
  \vspace*{-1cm}
  \caption{ The subprocess at the lowest order. }  \label{subpro}
}
\end{figure}

In this paper, to extract the polarized gluon distribution $\dg$,
we propose another process,
$p + \vec{p} \rightarrow \plmc +X \label{pro1}$
(Fig.~\ref{mainpro}),
which would be observed in the forthcoming RHIC experiment.
In this process, $\lmc$ is dominantly produced via fragmentation 
of a charm quark originated from gluon--gluon fusion.
The reason we focus on  this process is as follows.
The $\lmc$ is composed of a heavy quark $c$ and antisymmetrically
combined light $u$ and $d$ quarks.  
Hence, the $\lmc$ spin is basically carried by a charm quark
which is produced via gluon--gluon fusion at the lowest order in this
process as shown in Fig.~\ref{subpro}
\footnote{Since charm quarks are not main constituents of the proton,
the  gluon--gluon fusion process is dominant for charm quark production.}. 
Therefore, observation of the spin of the produced
$\lmc$ gives us information about the polarized gluons in the proton.
%
%
%
%
In order to study this attractive process;
\begin{equation}
p(p_{{}_A})+\vec{p}(p_{{}_B}) \rightarrow 
\plmc (p_{\lmc}) +X \label{pro2},
\end{equation}
whose subprocess at the lowest order is
\begin{equation}
g(p_{{}_a})+\vec{g}(p_{{}_b}) \rightarrow 
\vec{c}(p_c) + \bar{c}(p_{\bar{c}}) \label{subpro2},
\end{equation}
where $p_{{}_i}$ denotes the four-momenta of the $i$-particle
\footnote{For example, if $i=\lmc$, $p_{i}$ indicates the four-momenta of
$\lmc$.} and the over-arrow means that 
the particle polarization is either provided ($g$) or measured ($c$),
we introduce  two useful observables which will be measured by
the forthcoming RHIC experiment:
one is  the spin correlation
differential cross section, $d \Delta \sigma/ d\pt $,
and
the other is  the spin correlation asymmetry, $\all$.
They are defined as follows; 
\begin{eqnarray}
&&\frac{d \Delta\sigma}{d\pt}
\equiv \frac{
d\sigma(++) - d\sigma(+-) + d\sigma(--) - d\sigma(-+)}
{d\pt},\label{dsigma} \\
&&\all \equiv 
\frac{ \left[d\sigma(++) - d\sigma(+-) + d\sigma(--) - d\sigma(-+)\right]/d\pt}
     { \left[d\sigma(++) + d\sigma(+-) + d\sigma(--) + d\sigma(-+)\right]
/d\pt},\non\\
&&\phantom{\all}\equiv \frac{d \Delta \sigma/d\pt}{d \sigma/d\pt},
\label{ALL}
\end{eqnarray}
where $d\sigma(+-)/d\pt$, for example, denotes the spin-dependent differential
cross section with the positive helicitiy of the target proton and
the negative helicity of the produced $\lmc$.

Let us consider the process in the proton-proton c.m. frame 
and take the four-momenta of $p_{{}_i}$ as follows;
\begin{eqnarray}
&&p_{{}_{A,B}}=\frac{\sqrt{s}}{2}(1,\pm \beta ,\vec{0})\ \ {\rm with} \ \  
\beta \equiv \sqrt{1-\frac{4m^2_{p}}{s}}, \non \\
&&p_{\lmc}=(E_{\lmc},p_{{}_L},\vec{p}_{{}_T})\non \\
&& \phantom{p_{\lmc}}
=(\sqrt{m_{\lmc}^2 + \pt^2 \cosec^2 \Theta},
\pt\cot \Theta,\vec{p}_{{}_T}),\non\\
&&p_{{}_{a,b}}=x_{{}_{a,b}}p_{{}_{A,B}}\ \ ,\ \ p_{{}_c}=\frac{p_{\lmc}}{z},
\label{kine}
\end{eqnarray}
where the first, second and third components 
in parentheses are the energy,
the longitudinal momentum and the transverse momentum, respectively
\footnote{Here we do not  neglect the masses of all particles,
though we basically follows the method of Ref.~\cite{QCD}.}.
$\Theta$ and $m_{i}$ represent the c.m. scattering
angle of the produced $\lmc$ and mass of the $i$-particle,
respectively. 
$x_{a,b}$ and $z$ are the momentum fraction of the proton carried
by the gluon and the one of the charm quark carried by $\lmc$,
respectively.
Notice that we assume that the scattering angle of the $\lmc$ produced in 
the final state and the angle of the charm quark in the subprocess 
are almost same,
$\theta_{c}\simeq \theta_{\lmc}\equiv \Theta $.
This assumption is not unreasonable, because the momentum of $\lmc$  is
almost carried by a charm quark whose mass is much larger than the other
constituents of $\lmc$.
Then, using Eq.(\ref{kine}), we define the following Lorentz invariant 
variables;
\begin{eqnarray}
&&\stil \equiv s-2\mpr^2,\non \\
&&\ttil \equiv t-\mpr^2-\mlc^2
=-\sqrt{s}\left[
\sqrt{\mlc^2+\pt^2 \cosec^2 \Theta }+\beta \pt \cot \Theta 
\right],\non \\
&&\util \equiv u-\mpr^2-\mlc^2
=-\sqrt{s}\left[
\sqrt{\mlc^2+\pt^2 \cosec^2 \Theta}-\beta \pt \cot \Theta 
\right],
\label{mandel}
\end{eqnarray}
where $s$, $t$ and $u$ are conventional Mandelstam variables, 
$s=(\pA+\pB)^2$, $t=(\plc-\pB)^2$
and $u=(\plc-\pA)^2$, respectively.
Furthermore, for the subprocess, we define
\begin{equation}
\ths\equiv \hat{t}-\mc^2=\frac{x_b}{z}\ttil,\ \ \ 
\uhs\equiv \hat{u}-\mc^2=\frac{x_a}{z}\util, \label{mandel2}
\end{equation}
with the subprocess Mandelstam variables
$\hat{s}=(\pa+\pb)^2$, 
$\hat{t}=(\pc-\pb)^2$ and $\hat{u}=(\pc-\pa)^2$.

Using Eq.(\ref{kine})$\sim$Eq.(\ref{mandel2}),
the spin correlation differential cross section(Eq.(\ref{dsigma})) can
be expressed as
\begin{eqnarray}
&&\frac{d \Delta \sigma}{d \pt}
 = \int^{\Theta{{}^{\rm max}}}_{\Theta{{}^{\rm min}}}
\int^{1}_{x^{{}^{\rm min}}_{{}_a}} 
\int^{1}_{x^{{}^{\rm min}}_{{}_b}}
          G_{p_{{}_A}\rightarrow g_{{}_a}}(x_a,Q^2)
   \Delta G_{\vec{p}_{{}_B} \rightarrow \vec{g}_{{}_b}}(x_b,Q^2)
   \Delta {\rm D}_{\vec{c}\rightarrow \vec{\Lambda}_c^+}(z)\non \\
&&\phantom{\frac{d \Delta \sigma (\lambda,h)}{d \pt}=}
\times \frac{d \Delta \hat{\sigma}}{d \hat{t}}
J
dx_a dx_b d\Theta,
\label{dcross}
\end{eqnarray}
where $ G_{p_{{}_A} \rightarrow g_a}(x_a,Q^2)$,
$\Delta G_{\vec{p}_{{}_B} \rightarrow \vec{g}_b}(x_b,Q^2)$
and $\Delta {\rm D}_{\vec{c} \rightarrow \vec{\Lambda}_c^+}(z)$  
represent the unpolarized gluon distribution function, the polarized gluon
distribution function and the spin-dependent fragmentation function of the
outgoing charm quark decaying into a polarized $\plmc$, respectively.
Moreover $J$ is the Jacobian which transform the variables 
 $z$  and 
$\hat{t}$
into  $\Theta$ and $\pt$. 
Unfortunately, at present there is no established spin-dependent
fragmentation  functions because of lack of experimental data.
However, since a charm quark is much heavier than other constituents of
$\lmc$,
it is expected to be very rare for a charm quark to change  its spin
alignment during fragmentation process.
Therefore, it is not unreasonable to substitute 
${\rm D}_{c\rightarrow \lmc}$ for 
$\Delta{\rm D}_{\vec{c}\rightarrow \vec{\Lambda}_c^+}$.
In this work, we use the model by Peterson et al.~\cite{peter} 
for both ${\rm D}_{c\rightarrow \lmc}$ 
and  $\Delta{\rm D}_{\vec{c}\rightarrow \plmc}$
\footnote{
Actually, we used the Peterson functional form presented by
the Particle Data Group~\cite{pd}, which is normalized as
$\int {\rm D}_{c\rightarrow \lmc}(z) dz =0.503.$}.
For the subprocess, the spin correlation differential cross section,
$d \Delta \hat{\sigma}/d \hat{t}$
is calculated to be
\begin{eqnarray}
&&\frac{d \Delta\hat{\sigma}}{d \hat{t}}
=\frac{\pi \alpha^2_s}{\shs}
\left[ \frac{\mc^2}{24} \left\{\frac{9 \ths -19 \uhs}{\ths \uhs}
+\frac{8\shs}{\uhs^2}\right\}
+\frac{\shs}{6}\left\{\frac{\ths-\uhs}{\ths \uhs}\right\}
-\frac{3}{8}\left\{ \frac{2 \ths}{\shs} +1 \right\}
\right]
\label{cross}
\end{eqnarray}
and the Jacobian,   
$J$,
is given by
\begin{equation}
J=\frac{2s\beta \pt^2 \cosec^2 \Theta}
{z\stil \sqrt{m^2_{\lmc}+\pt^2 \cosec^2 \Theta}}.
\label{jabocian}
\end{equation}
where $z$ is
\begin{equation}
z=\frac{x_1}{x_a}+\frac{x_2}{x_b}
\end{equation}
with $x_1=-{\ttil}/{\stil}$ and $x_2=-{\util}/{\stil}$.
The minimum of $x_a$, $x_b$ are given by 
\begin{equation}
x^{{}^{\rm min}}_{{}_a}=\frac{x_1}{1-x_2},\ \  
x^{{}^{\rm min}}_{{}_b}=\frac{x_a x_2}{x_a-x_1}.
\label{jacobi}
\end{equation}
The unpolarized differential cross sections were calculated by Babcock
et al. \cite{ggcc}.

For numerical calculation, we use as input parameters,
$m_c = 1.5$ GeV, $m_p = 0.938$ GeV and $\mlc = 2.28$ GeV~\cite{pd}.
We limit the integration region of $\Theta$ and 
$\pt$ of produced $\lmc$  as
$\frac{\pi}{6} \leq \Theta \leq \frac{5\pi}{6}$ and
3 GeV $\leq \pt \leq$ 20 GeV for  $\sqrt{s}=200$ GeV and 500 GeV,
in order to get rid of  the contribution of
the diffractive $\lmc$ production and also the $\lmc$ production 
through a single charm quark production via $W$ boson exchange 
and $W$ boson production. 
As for the gluon distributions, we take the GS96(set-A and -B)~\cite{gs} 
and the GRSV96~\cite{grsv} parameterization models for the polarized
gluon distribution function and the GRV95~\cite{grv} model for 
the unpolarized one.
Note that the GS96 and GRSV96 models can excellently
reproduce experimental results for
the polarized structure function of nucleons, 
though behavior of these polarized gluon distributions is quite different.
In other words, the data on  polarized structure functions 
of nucleons and deuteron alone are not enough to distinguish the model 
of gluon distribution functions.
We are interested in the sensitivity of those observables of 
Eq.(\ref{dsigma}) and Eq.(\ref{ALL}) on 
the polarized gluon distributions in the nucleon 
in this process.    
%
%
\begin{figure}[t!]
  \begin{center}
    \epsfxsize=7cm
    \epsfbox{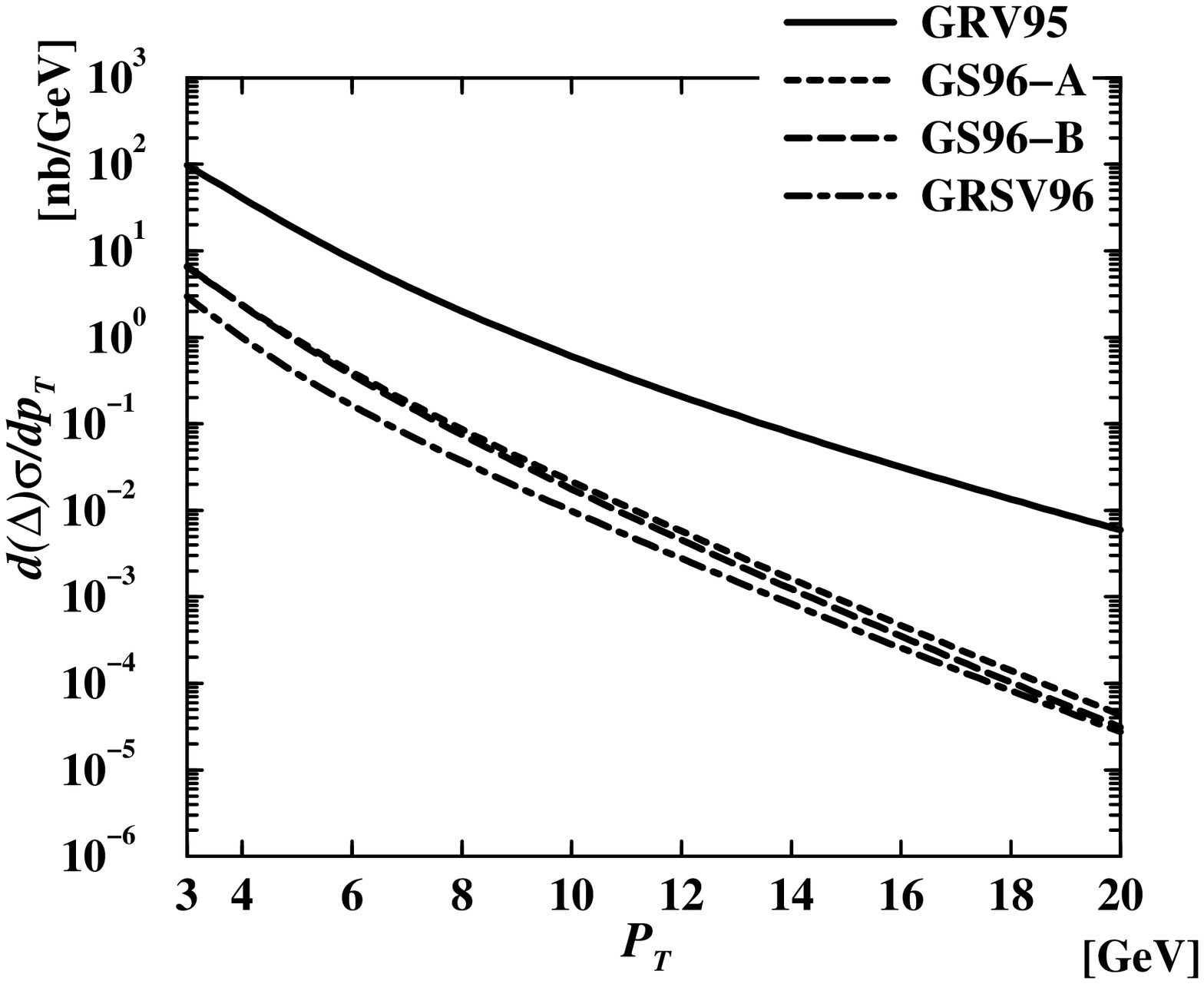}
   \epsfxsize=7cm
    \epsfbox{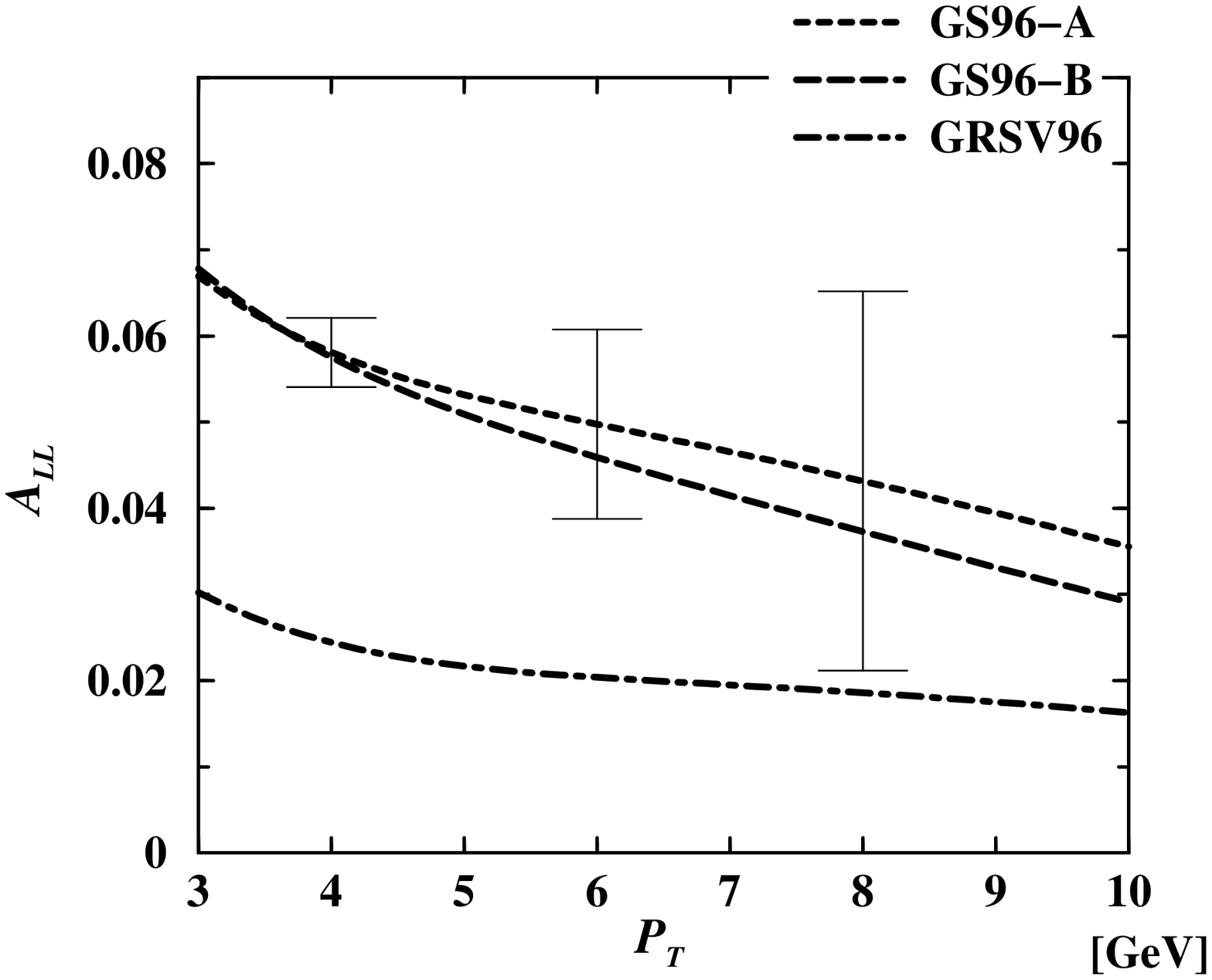}
  \end{center}
  \vspace*{-1cm}
  \caption{The unpolarized and spin correlation differential cross
 section (left panel) and the spin correlation asymmetry (right panel) 
as a function of 
$\pt$ at $\sqrt{s}=200$ GeV.
The solid line in the left figure represents the unpolarized
differential cross section with the GRV95 model for the unpolarized
gluon distribution.
The dashed, long-dashed and  dot-dashed lines indicate 
numerical results with the set A, B of the GS96 model 
and the GRSV96 model, respectively, for the polarized gluon 
distribution.
The error bars for the dashed line shows the statistical sensitivity
(see text).
} \label{200fig}
  \begin{center}
    \epsfxsize=7cm
    \epsfbox{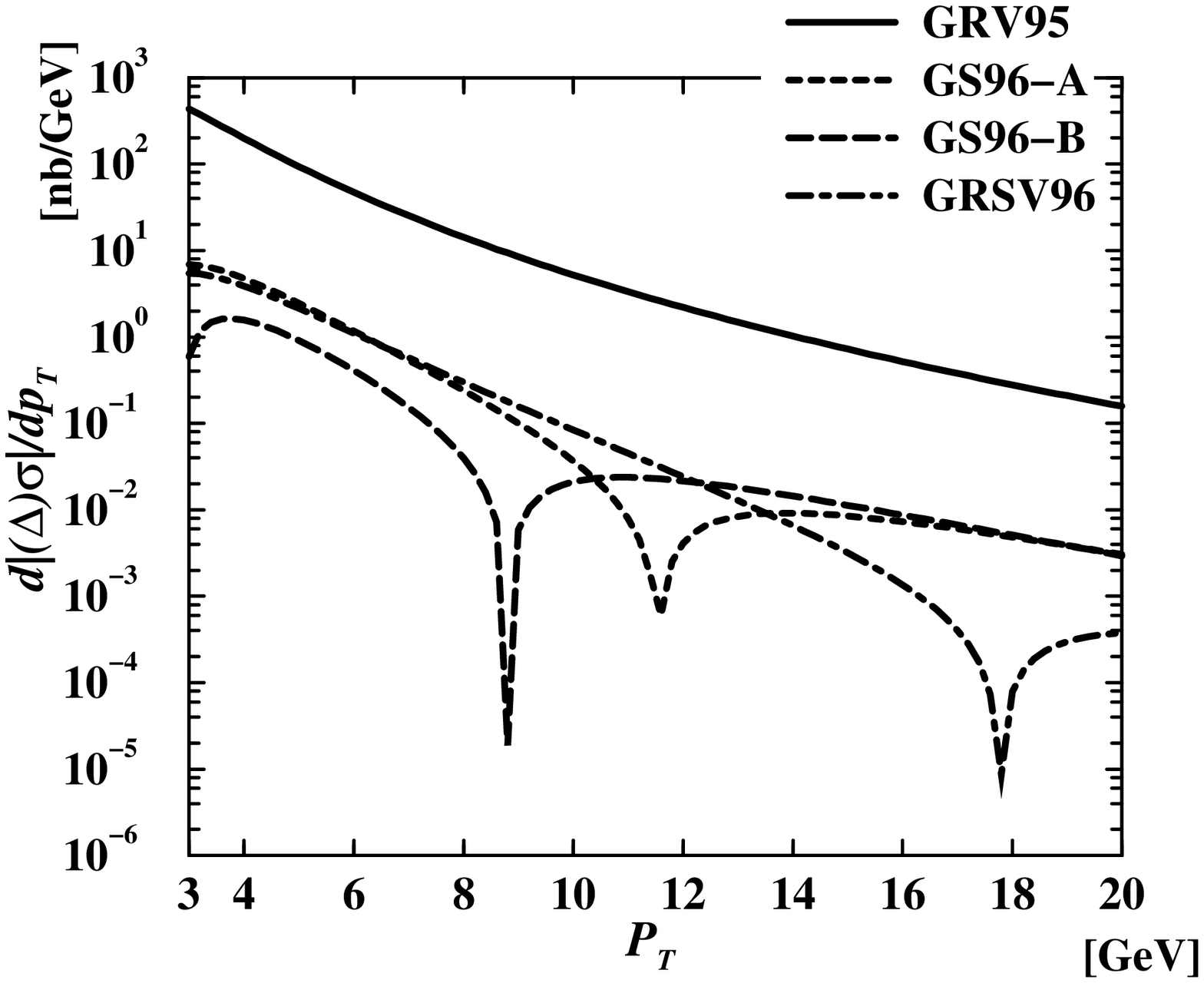}
   \epsfxsize=7cm
    \epsfbox{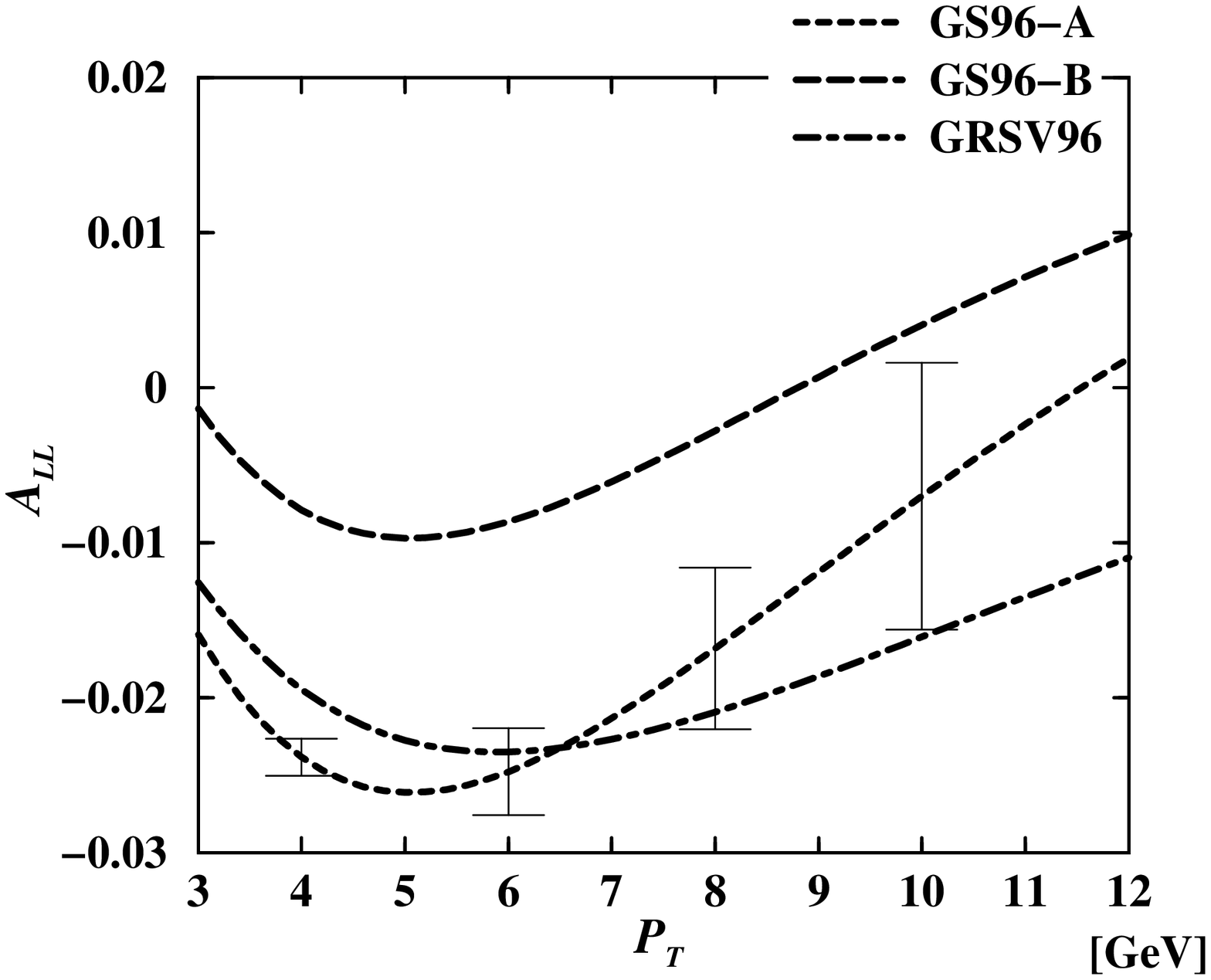}
  \end{center}
  \vspace*{-1cm}
  \caption{The same as in Fig. \ref{200fig}, but for $\sqrt{s}=500$ GeV.}
\label{500fig}
\end{figure}
%

We show the $\pt$ distribution of $d \Delta \sigma /d\pt$ and 
$\all$ in Fig.~\ref{200fig} for $\sqrt{s} = 200$ GeV  
and in Fig.~\ref{500fig} for
$\sqrt{s} = 500$ GeV, respectively.
Notice that in Fig.~\ref{500fig} the absolute value of 
$d \Delta \sigma /d \pt$ is presented,
since the negative value of $d \Delta \sigma /d\pt$ cannot be depicted
in the figure which has an ordinate with logarithmic scale.
Actually, for the case of $\sqrt{s}=500$ GeV, the value of 
$d \Delta \sigma /d\pt$ becomes negative for $\pt$ smaller than
the value corresponding to the sharp dip shown in Fig.~\ref{500fig}.
On the other hand, $d \Delta \sigma /d \pt $ at $\sqrt{s}=200$ GeV 
is positive for all $\pt$ regions as shown in  Fig.~\ref{200fig}.
To understand this quite different behavior
of $d \Delta \sigma /d\pt $  depending on $\sqrt{s}$,
some comments are in order.
As seen from Eq.(\ref{dcross})$\sim$ Eq.(\ref{jacobi}),
the  sign of $d \Delta \sigma /d\pt$ is determined by the sign of
$d\Delta \hat{\sigma}/d \hat{t}$
because all variables except for $d\Delta \hat{\sigma}/d \hat{t}$ 
are positive in whole kinematical regions of $x_a$, $x_b$ and $\Theta$.
Since  both $\ths$ and $\uhs$ are negative and smaller than $\shs$ in
magnitude,
the sign of $d \Delta \hat{\sigma} /d \hat{t}$  mainly originates from
the second term of the right hand side of Eq.(\ref{cross}), in which
 $\ths - \uhs$ is calculated to be
\begin{equation}
\ths-\uhs \simeq x_b\sqrt{s}
\left\{
x_a \sqrt{s}-\frac{2}{z}
\left( \sqrt{\mlc^2+\pt^2 \cosec^2 \Theta}+ \pt \cot \Theta 
\right)
\right\},
\label{t-u}
\end{equation}
in the limit of the massless proton
\footnote{In the region of $s$ much larger than $m_p^2$ where we are now
focusing, this limit with $s=\stil$ and $\beta=1$ is 
a good  approximation.}.
Furthermore, in the same limit,  $x_a^{\rm min}$ reduces to 
\begin{equation}
x_a^{\rm min} \simeq 
\frac{\sqrt{\mlc^2+\pt^2 \cosec^2 \Theta} +\pt \cot \Theta}
{\sqrt{s}- \left( \sqrt{\mlc^2+\pt^2 \cosec^2 \Theta}- \pt \cot \Theta
\right)}
\label{xa}
\end{equation}
from Eq.(\ref{jacobi}).
Eq.(\ref{xa}) shows that $x_a$ becomes very small in some region of
$\pt$ and $\Theta$ 
when  $\sqrt{s}$ is large and hence leads to large values of the gluon
distributions,  $G_{p_{{}_A}\rightarrow g_{{}_a}}(x_a,Q^2)$.
Thus when Eq.(\ref{t-u}) becomes negative due to very small $x_a$,
the negative contribution of $d \Delta \hat{\sigma} /d\hat{t}$ to
 $d \Delta \sigma / d\pt$ becomes larger for $\sqrt{s}=500$ GeV
than for $\sqrt{s}=200$ GeV 
because $G_{p_{{}_A}\rightarrow g_{{}_a}}(x_a,Q^2)$ is
larger at smaller $x_a$.
Therefore, for large $\sqrt{s}$ such as $\sqrt{s}=500$ GeV,
even after integration,
$d \Delta \sigma / d\pt$ is negative 
for $\pt$ smaller than
the value corresponding to the dip, as shown in  Fig.~\ref{500fig}.
It is very interesting to note that the behavior of $\all$ strongly
depends on $\sqrt{s}$ as shown from Figs.~\ref{200fig} and~\ref{500fig}.
In any case, both figures show that $\all$ is sensitive to 
the polarized gluon distribution function. 

Finally, to examine if our predictions can be tested
at the forthcoming RHIC experiment, 
we estimate the statistical sensitivity of 
the spin correlation asymmetry, $\delta \all$, 
according to the  method given in Ref\cite{erra}.
The value of $\delta \all$ is estimated by
\begin{equation}
\delta \all \simeq \frac{1}{P}
\frac{1}{\sqrt{ b_{\lmc}\   \varepsilon\  {\cal L} \ T \ \sigma}}.
\label{err}
\end{equation}
We can estimate the statistical sensitivity, $\delta \all$,
for $T =$100-day experiments 
by using the parameters of the beam polarization ($P=70\%$), 
a luminosity (${\cal L} =8 \times 10^{31}~(2 \times 10^{32})
{~\rm cm^{-2} sec^{-1}}$ for  $\sqrt{s}=200~(500)$ GeV),
the trigger efficiency ($\varepsilon \equiv 10 \%$ )
of detecting high $\pt$ charm production events
through their semi-leptonic decays
and a branching ratio 
($ b_{ \lmc } \equiv Br( \lmc \rightarrow p K^- \pi^+ ) \simeq 5 \% $
\cite{pd})
\footnote{Here, we simply assumed the efficiency for reconstructing
the $p K^- \pi^+$ decay of $\lmc$ to be 100 $\%$ by quoting
only the decay branching fraction.
Although the actual efficiency might be more likely to be less than
10 - 20 $\%$ rather than more than 50 $\%$ 
(almost perfect, background free measurement),
this should be studied by experimentalists 
in the forthcoming RHIC experiment.}.
$\sigma$ denotes the unpolarized
cross section integrated over suitable  $\pt$ region.
The results are summarized  in the right panels of
Figs.~\ref{200fig} and~\ref{500fig}.
As shown in these figures,
$\delta \all$ is smaller than the difference of the model predictions
at moderate $\pt$ region and thus our predictions are  expected to be
tested in the RHIC experiment, though 
the differential cross section, $d \sigma / d\pt $, becomes 
rapidly smaller  with increasing $\pt$, 
and thus in the larger $\pt$ region,
$\delta \all$
becomes too large to distinguish the model of polarized gluon 
distribution functions.
Furthermore, note that $\delta \all$ is expected to
become small  if
a trigger efficiency $\varepsilon $ is improved and/or 
another decay modes of $\lmc$ are additionally taken into account.
Moreover, the statistics should be twice
by using the $\overline{\Lambda^{{}^+}_c}$ data as well,
since high $\pt$ charm production is dominated by gluon fusion,
as long as the $\lmc$ decays satisfy $CP$ invariance.

In summary, to extract the polarized gluon distribution 
$\Delta G(x,Q^2)$,
we proposed a new polarized process,
$p+\vec{p}\rightarrow\plmc+X$,
which could be observed in the  forthcoming RHIC experiment.
We calculated the the spin correlation differential cross section,
$d \Delta \sigma/d \pt$,
and the spin correlation asymmetry, $\all$,
and found that $\all$ is quite sensitive to the polarized gluon
distribution functions; we can rather clearly
distinguish the model of polarized gluon distribution functions
at moderate $\pt$ region.
It is also remarkable that the $\all$ shows very different behavior in
the region of $\sqrt{s}=200 \sim 500$ GeV, covered by RHIC. 
Therefore, the process looks 
promising for testing the model of polarized gluon 
distribution functions,
though the present analysis is based on the leading
order calculation.
In order to get  more profound information on the  behavior of polarized
gluons, we need the next-to-leading order calculation. 
Furthermore, we also need further investigation of the polarized
fragmentation functions,
$\Delta {\rm D}_{\vec{c} \rightarrow \vec{\Lambda}_c^+}(z)$,  
to get more reliable prediction.
Although these subjects are interesting and important in their own
right, they are out of scope in this work.

We hope our prediction will  be tested in the forthcoming RHIC
experiment. 
\\
\\
One of the authors (T.M.) would like to thank for the financial
support by the Grant-in-Aid for Scientific Research,
Ministry of Education, Science and Culture, Japan (No.11694081).
\vskip 0.5cm

\end{document}